\documentstyle [12pt]{article}
\begin{document}
%\psnoisy
%\psfigurepath{figs}
%\def\ps{post\-script}
\title{ $\rho N$ resonance dynamics in the proton nucleus reaction }
\author
{ Swapan Das  \\
Nuclear Physics Division, Bhabha Atomic Research Centre, \\
Mumbai-400 085, India }
%%%\date{\today}%
\maketitle

\begin{abstract}
The strong coupling of rho meson to the nucleon produces $s$ and $p$
wave rho meson nucleon $(\rho N)$ resonances. In a nucleus, the $\rho N$
resonance-hole polarization generates the optical potential or
self-energy for the rho meson. The scattering of rho meson due to this
potential provides valuable informations about the $\rho N$ resonance
dynamics in a nucleus.
To investigate it, we use this potential to calculate the mass distribution
spectrum for the $\rho$ meson produced coherently in the proton nucleus
reaction. The cross sections arising due to $s$ and $p$ wave $\rho N$
resonances have been presented. The coherent and incoherent contributions to
the cross sections due to these resonances are compared. In addition, the
calculated results due to non-relativistic and relativistic $\rho$ meson
self-energy are illustrated.
\end{abstract}
\medskip

PACS number(s): 25.40.Ve, 24.30.-v, 24.50.+g

{\it Key words}:
Proton-nucleus reaction, $\rho$ meson-nucleon resonance dynamics,
$\rho$ meson production

\section{ INTRODUCTION }

The rho meson nucleon resonances arise because of the strong coupling
of this meson to the nucleon. The importance of the hadronic resonances
in the particle production phenomena is undoubtedly understood from
long ago. For example, the pion production data in the nuclear reaction
below 1 GeV had been interpreted successfully by the formation of the
$\Delta (1232)$ resonance in the intermediate state \cite{delta}. Therefore,
the dynamics of the $\rho N$ resonances can have intimate
relationship with the rho meson production in the nuclear reaction.
The sub-threshold rho meson emission is well described
by the $N(1520)$ resonance production in the intermediate state \cite{sbrh}.
At higher energy, the importance of $N(1720)$ resonance has been discussed
\cite{dasr} in context to the $\rho$ meson production in the proton nucleus
reaction. Therefore, the study of the rho meson production in the nuclear
reaction can unveil many interesting physics of the $\rho N$ resonance
dynamics in a nucleus.

The $\rho N$ resonances were also found significant to investigate the
modification of the rho meson in the nuclear medium. Sometime back,
Kondratyuk et al., \cite{kscge} estimated the rho
meson nucleon scattering amplitude, i.e., $f_{\rho N}$, due to the
formation of $\rho N$ resonances in the intermediate state. Using this
amplitude, they have shown the $\rho$ meson mass in a nucleus is below
770 MeV in the static limit. This agrees with the result of scaling
hypothesis formulated by Hatsuda and Lee \cite{htle}.
Friman et al., \cite{frpr} calculated the $\rho$ meson self-energy arising
due to $p$ wave $\rho N$ scattering, via $N(1720)$ and $\Delta(1905)$
resonances. According to their calculation, this self-energy (which appears
in the $\rho$ meson spectral function) reduces the mass of the $\rho$ meson
at high baryon density.
Peters et al., \cite{ppllm} extended this calculation by incorporating
contributions from all four starred $s$ and $p$ wave $\rho N$ resonances.
As shown by them, the $s$ wave $\rho N$ resonances (specifically, $N(1520)$)
have significant influence on the $\rho$ meson spectral function.
The $\rho$ meson self-energy due to $s$ wave $\rho N$
resonance-hole polarization shows an important feature, as it depends upon
both energy and momentum of the rho meson. Therefore, the self-energy due to
$s$ wave $\rho N$ resonance can contribute to the $\rho$ meson spectral
function in the static limit. This is unlike to that occurring due to $p$
wave $\rho N$ scattering, where the rho meson self-energy depends only on
its momentum.

The self-energy or optical potential is inevitable to describe the elastic
scattering of a particle by the nucleus. Using it, we studied various
aspects for the coherent (elastic) scattering of $\rho^0$ meson in the
proton nucleus reaction \cite{dscr}. In one part of this study, we
investigated the sensitivity of the coherent $\rho^0$ meson mass
distribution on its optical potential formulated by various authors
\cite{kscge,htle,frpr,brrh} including Peters
et al., \cite{ppllm}.
The later authors, as mentioned above, have shown non-relativistically that
various $\rho N$ resonances can generate the self-energy for the rho
meson in a nucleus. Due to lack of scope, we could not show in our
earlier work \cite{dscr} the distinct contribution arising from each
$\rho N$ resonance to the $\rho^0$ meson production cross section.
Therefore, those are illustrated
in the present work. In addition, we extend our calculation to accommodate
the relativistic evaluation of the $\rho$ meson self-energy due to these
resonances and discuss how the $\rho N$ resonance behave relativistically
in the above reaction.
We compare the non-relativistic and relativistic optical potentials for the
$\rho^0$ meson arising due to each $\rho N$ resonance-hole polarization in
the nucleus. We also present the results for the coherent and incoherent
contributions of these resonances to the overall cross sections.

\section{ FORMALISM }

The coherent $\rho^0$ meson production in the proton nucleus reaction
describes the elastic scattering of the virtual rho meson (emitted by the
beam proton) to its real state. Symbolically, this reaction is used to
express as
$ p + A \mbox{(g.s.)} \to p^\prime + \rho^0 + A \mbox{(g.s.)} $.
As we worked out earlier \cite{dscr}, the double differential cross section
for the coherent $\rho$ meson mass distribution in the $(p,p^\prime)$
reaction on a nucleus is given by
\begin{equation}
\frac{d\sigma}{dmd\Omega_{p^\prime}} =
\int \int d\Omega_\rho dE_{p^\prime} [KF] S(m^2) <|T_{fi}|^2>,
\label{dsgm1}
\end{equation}
where $[KF]$ is the kinematical factor associated with this reaction. The
expression for it is
\begin{equation}
[KF] =  \frac{ m^2_p E_{A^\prime} }{ (2\pi)^5}
\frac{ k_{p^\prime} k^2_{\rho} m }
{ k_p | k_\rho ( E_i - E_{p^\prime} ) - ( {\bf k}_p - {\bf k}_{p^\prime} )
\cdot {\hat k}_\rho  E_\rho | }.
\label{ekf}
\end{equation}

The factor $S(m^2)$ in Eq.~(\ref{dsgm1}) denotes the mass distribution
function for the rho meson of mass $m$ in the free space. It is well
addressed by the Breit-Wigner form \cite{elena}:
\begin{equation}
S(m^2) = \frac{ 1 }{ \pi } \frac{ m_\rho \Gamma_\rho (m^2) }
{ [ ( m^2 - m_\rho^2 )^2 + m^2_\rho \Gamma^2_\rho (m^2) ] },
\label{mdfn}
\end{equation}
with $ m_\rho (\simeq 770 ~\mbox{MeV}) $ being the resonance mass of the
$\rho$ meson in the free state. $ \Gamma_\rho (m^2) $ represents the total
free space decay width for the $\rho$ meson of mass $m$,
arising dominantly $(\sim 100\%)$ due to $ \rho^0 \to \pi^+\pi^- $.
Therefore, $ \Gamma_\rho (m^2) $ can be written as
\begin{equation}
\Gamma_\rho (m^2) \approx
\Gamma ( m^2_\rho )_{\rho \to \pi \pi} \biggl ( \frac{ m_\rho }{ m } \biggr )
\biggl [ \frac{ k( m^2 ) }{ k(m^2_\rho) } \biggr ]^3 \Theta ( m^2 - 4 m^2_\pi ),
\label{rhwd}
\end{equation}
with $ \Gamma (m^2_\rho) _ {\rho \to \pi\pi} \approx 150 $ MeV. $ k (m^2) $
is the pion momentum in the rest frame of decaying $\rho^0$ meson of mass
$m$. This expression intrinsically implies the detection of rho meson
through its decay products $\pi^+\pi^-$ in the final state. We have not
considered here the distortion due to pion nucleus scattering since the
effect of it (in this kind of reaction) has been found insignificant
\cite{dscr2}.

$T_{fi}$ in Eq.~(\ref{dsgm1}) is the $T$-matrix for the coherent rho meson
production in the proton nucleus reaction. The annular bracket around
$|T_{fi}|^2$ represents the average and summation over spins and
polarization of the initial and final states respectively.  $T_{fi}$ is
given by
\begin{eqnarray}
T_{fi} = \int \int d{\bf r^\prime} d{\bf r}  ~
\chi^{(-)*} ({\bf  k}_\rho,{\bf r^\prime}) \Pi_\rho ({\bf r^\prime})
G_\rho ({\bf r^\prime - r}) \Psi_\rho ( {\bf q}, {\bf r}).
\label{tmx0}
\end{eqnarray}
In this equation, $ G_\rho ({\bf r^\prime - r}) $ describes the propagation
of the virtual $\rho^0$ meson from ${\bf r}$ to ${\bf r^\prime}$, where
the elastic scattering of this meson (to its real state) is taken place
due to the $\rho$ meson nucleus optical potential
$ V_{O\rho} ({\bf r^\prime}) $. This potential appears in the above equation
through the $\rho$ meson self-energy, i.e., $ \Pi_\rho
({\bf r^\prime}) = 2 q_0 V_{O\rho} ({\bf r^\prime}) $. Here, $ q_0
( = E_p-E_{p^\prime} ) $ is the energy of the virtual rho meson. $ \Pi_\rho
({\bf r^\prime}) $ will be elaborated in the next section.

The symbol $ \Psi_\rho ( {\bf q}, {\bf r}) $ in Eq.~(\ref{tmx0}) stands for
the virtual rho meson production amplitude in the $(p,p^\prime)$ reaction.
The expression for it is
\begin{equation}
\Psi_\rho ( {\bf k}_{p^\prime}, {\bf k}_p; {\bf r})
= \chi^{(-)*} ({\bf k}_{p^\prime},{\bf r} )
\Gamma_{\rho NN} \chi^{(+)} ({\bf k}_p,{\bf r}),
\label{vrp}
\end{equation}
where $\Gamma_{\rho NN}$ denotes the vertex function at the
$\rho^0 pp^\prime$ vertex. It is governed by the $\rho NN$ Lagrangian as
described in Ref.~\cite{dscr}. $\chi$s represent the distorted wave
functions for protons in the continuum. In the energy region of $\rho$
meson production, the distortion arising due to proton nucleus scattering
is purely absorptive. Therefore, the distorted wave functions
for protons can be approximated by their plane waves multiplied by the
attenuation factor $ {\cal A_F} $ for them \cite{anfc}, i.e.,
\begin{equation}
\chi^{(-)*} ({\bf k}_{p^\prime},{\bf r} ) \chi^{(+)} ({\bf k}_p,{\bf r})
\approx  {\cal A_F}  e^{i{\bf q}.{\bf r}};
~~~~~ {\bf q} = {\bf k}_p - {\bf k}_{p^\prime}.
\label{dwfnp}
\end{equation}
This equation shows ${\bf q}$ is the momentum of the virtual rho meson
emitted at the $\rho NN$ vertex.

The $\rho^0$ meson in the final state can emit in all directions. Therefore,
the partial wave expansion is preferred to express its wave function
$ \chi^{(-)} ({\bf  k}_\rho,{\bf r^\prime}) $ appearing in
Eq.~(\ref{tmx0}), i.e.,
\begin{equation}
\chi^{(-)*} ( {\bf k_\rho, r} )
= \frac{4\pi}{k_\rho r} \sum_{lm} (-i)^l u_l (k_\rho r)
                        Y^*_{lm}(\hat{k_\rho}) Y_{lm}(\hat{r}).
\label{dwrh}
\end{equation}
The radial part of this wave function, i.e., $u_l (k_\rho r)$, is generated
by solving the relativistic wave (Schr$ \ddot{\mbox{o}} $dinger) equation
using the $\rho$ meson optical potential $ V_{O\rho} ({\bf r}) $,
described later.

\section{ RESULTS AND DISCUSSION }

The attenuation factor ${\cal A_F}$ in Eq.~(\ref{dwfnp}), which accounts
the distortion for  protons, can be estimated by using the following
equation \cite{anfc}:
\begin{equation}
{\cal A_F} = \frac{1}{A} \int d{\bf b} T({\bf b}) exp[ -Im\delta({\bf b}) ],
\label{attn}
\end{equation}
where $A$ is the mass number of the nucleus.
$ T({\bf b}) [ = \int ^{+\infty} _{-\infty} \varrho ({\bf b},z) dz ] $
represents the thickness function for the nucleus at the impact parameter
{\bf b}. $ \varrho ({\bf b},z) $ describes the density distribution for the
nucleus, given in Eq.~(\ref{dc12}) for $^{12}$C nucleus. $\delta ({\bf b})$
denotes the total phase shift function for the proton nucleus scattering,
i.e., $ \delta ({\bf b})
= \delta_p ({\bf b}) + \delta_{p^\prime} ({\bf b}) $. The form for it is
given by
\begin{equation}
\delta_p ({\bf b})
= - \frac{1}{v_p} \int ^\infty_0  V_{Op} ({\bf b}, z) dz,
\label{phsf}
\end{equation}
where $v_p$ is the velocity of the proton. $ V_{Op} ({\bf b},z) $ denotes
the optical potential for the proton nucleus scattering.

The imaginary part of the proton nucleus optical potential
$ V_{Op} ({\bf b},z) $ is required to evaluate the attenuation factor
${\cal A_F}$ for protons (see Eqs.~(\ref{attn}) and (\ref{phsf})). Using
the high energy ansatz, i.e., $``t\varrho"$ approximation, the imaginary
part of $ V_{Op} ({\bf b},z) $ can be written as
\begin{equation}
Im V_{Op} ({\bf r})
= -\frac{1}{2} v \sigma^{pN}_t \varrho ({\bf r}).
\label{oprh}
\end{equation}
$v$ is the relative velocity in the proton nucleon cm system.
$\sigma^{pN}_t$ denotes the total cross section for the proton nucleon
scattering. The energy dependent values for it is taken from the
measurements
\cite{nnsc,pdg}. $ \varrho ({\bf r}) $ describes the spatial shape of the
optical potential $ V_{Op} ({\bf r}) $. It is usually approximated by the
density distribution of the nucleus. The form of $ \varrho ({\bf r}) $ for
the $^{12}\mbox{C}$  nucleus, as extracted from the electron scattering
data \cite{adnd}, is given by
\begin{equation}
\varrho ({\bf r}) = \varrho_0 [ 1 + w(r/c)^2 ] e^{ -(r/c)^2 };
~~ w=1.247,  ~ c=1.649 ~ \mbox{fm}.
\label{dc12}
\end{equation}
This density is normalized to the mass numbers of the nucleus.

As mentioned earlier, we look for the contribution of the individual
$\rho N$ resonance to the cross section for the coherent rho meson
production in the proton nucleus reaction. Upto $\sim 1.9$ GeV, the
existence of these resonances (which emit $\rho$ meson) is confirmed
\cite{pdg}.
The formation of these resonances in a nucleus generates the $\rho N$
resonance-hole polarizations in it. The interaction energy associated with
these polarizations is referred as the self-energy or optical
potential for the rho meson. This potential can modify the $\rho$ meson
propagation through the nucleus.
Therefore, the $\rho$ meson production process in the nuclear reaction
can be thought as a potential tool to investigate the $\rho N$ resonance
dynamics in a nucleus.

To disentangle the above issue explicitly, we use the rho meson
self-energy (or optical potential) evaluated by Peter et al.,
\cite{ppllm}. According to them, the $\rho$ meson self-energy due to $s$
wave (negative parity) $\rho N$ resonance is given by
\begin{equation}
\Pi^{\mu\nu}_{(s)} (k_0, {\bf k})
= -(k^2_0 P^{\mu\nu}_T - k^2 P^{\mu\nu}_L)
    \left ( \frac{f_{RN\rho}}{m_\rho} \right )^2 F^\prime ({\bf k}^2)
    S_\Sigma \beta (k_0, {\bf k}),
\label{ses}
\end{equation}
with $k^2 = k^2_0-{\bf k}^2$. $k_0$ and ${\bf k}$ are the energy and
momentum respectively of the rho meson. $ F^\prime ({\bf k}^2) $ is
the monopole form factor associated with the $\rho$ meson self-energy:
$ F^\prime ({\bf k}^2) = \frac { \Lambda^{\prime 2} }
{ \Lambda^{\prime 2} + {\bf k}^2} $ with $ \Lambda^\prime = 1.5 $ GeV
\cite{ppllm}.
$P^{\mu\nu}_T$ and $P^{\mu\nu}_L$ denote the respective longitudinal and
transverse projection operators for the self-energy. They are normalized
as $ P^{\mu\nu}_L P_{\mu\nu L} = 1 $ and $ P^{\mu\nu}_T P_{\mu\nu T} = 2 $.
The longitudinal and transverse self-energies can be obtained by using
following relations \cite{ppllm}:
\begin{equation}
\Pi^{(s)}_L (k_0, {\bf k}) = -P^{\mu\nu}_L \Pi_{\mu\nu} (k_0, {\bf k});
~~~~~
\Pi^{(s)}_T (k_0, {\bf k})
= -\frac{1}{2} P^{\mu\nu}_T \Pi_{\mu\nu} (k_0, {\bf k}).
\label{pjop}
\end{equation}
The average $\rho$ meson self-energy due to $s$ wave $\rho N$ resonances
is given by $ \Pi^{(s)} (k_0, {\bf k}) =
\frac{1}{3} [ 2\Pi^{(s)}_T (k_0, {\bf k}) + \Pi^{(s)}_L (k_0, {\bf k}) ] $.
On the other hand, the $\rho$ meson self-energy due to $p$ wave (positive
parity) $\rho N$ resonance is purely transverse. The expression for it
\cite{ppllm} is
\begin{eqnarray}
\Pi^{(p)} (k_0, {\bf k}) \equiv \Pi^{(p)}_T (k_0, {\bf k})
= {\bf k}^2 \left ( \frac{f_{RN\rho}}{m_\rho} \right )^2 F^\prime ({\bf k}^2)
    S_\Sigma \beta (k_0, {\bf k}).
\label{sep}
\end{eqnarray}

In Eqs.~(\ref{ses}) and (\ref{sep}), $f_{RN\rho}$ denotes the constant for
the rho meson-nucleon coupling to the resonance $R$, and $S_\Sigma$
represents the associated spin-isospin transition factor. As shown by
Peters et al. \cite{ppllm}, there are four $s$-wave $\rho N$ resonances,
namely $N(1520)$, $\Delta (1620)$, $N(1650)$, $\Delta (1700)$, and five
$p$-wave $\rho N$ states $N(940)$, $\Delta (1232)$, $N(1680)$,
$N(1720)$ and $\Delta (1905)$. Values of $f_{RN\rho}$ and $S_\Sigma$ for
these resonances, as estimated by Peters et al. \cite{ppllm}, are listed
in the Table-1. For $\Delta (1232)$ resonance, the Landau-Migdal parameter
$g$ is introduced to account for the short-range correlation:
$ \Pi_\Delta^{(p)} (k_0, {\bf k})  \rightarrow
 \Pi_\Delta^{(p)} / \left [ 1 - g\frac{ \Pi_\Delta^{(p)} }{ k^2 } \right ] $,
with $g=0.5$. The inclusion of Landau-Migdal parameter for other
resonances was not recommended \cite{ppllm}. The resonance $N(1440)$ was
not incorporated in this calculation, since its width decaying to
the $\rho N$ channel is insignificant $(8<\%)$.

The Linhard function $ \beta (k_0, {\bf k}) $ appearing in Eqs.~(\ref{ses})
and (\ref{sep}) is given by
\begin{equation}
\beta (k_0, {\bf k})
= -\int^{p_F}_0 \frac{ d\bf{p} }{ (2\pi)^3 }
\left [ \frac{ 1 }{ k_0 -E_N({\bf p}) + E_R({\bf p+k}) }
      + \frac{ 1 }{-k_0 -E_N({\bf p}) + E_R({\bf p-k}) } \right ].
\label{lhfm}
\end{equation}
This integration eventually means the summation over all nucleons, i.e.,
$ \Sigma_i \to 4V\int^{p_F}_{0} \frac{ d{\bf p} }{ (2\pi)^3 } $.
$E_N ({\bf Q})$ and $E_R({\bf Q})$ represent energies for the nucleon
$N$ and resonance $R$ respectively. They are defined as
\begin{equation}
E_N({\bf Q})
= \sqrt{m^2_N+{\bf Q}^2};
~~~~~
E_R({\bf Q})
= \sqrt{m^2_R+{\bf Q}^2} - \frac{i}{2}\Gamma_R,
\label{enr}
\end{equation}
where $\Gamma_R$ denotes the width of the resonance $R$. It is composed
of the pion-nucleon and rho meson-nucleon decay widths, i.e.,
$ \Gamma_R = \Gamma_{ R \to N\pi } + \Gamma_{ R \to N\rho } $,
elaborated in the Ref.~\cite{ppllm}.

In the non-relativistic calculation done by Peters et al. ~\cite{ppllm},
the $\rho$ meson self-energies (given in Eqs.~(\ref{ses}) - (\ref{sep}))
were evaluated in the nucleon rest frame except the coupling constant
$f_{RN\rho}$ appearing in it. This constant was calculated in the
resonance rest frame. In the nucleon rest frame, the Linhard function
$ \beta (k_0, {\bf k}) $ in Eq.~(\ref{lhfm}) is reduced to
\begin{equation}
\beta (k_0, {\bf k})
= \frac{1}{2} \varrho
  \frac{ E_R({\bf k}) - m_N }{ k^2_0 - [E_R({\bf k}) - m_N]^2 }.
\label{bta}
\end{equation}
In this case, $k$ is the four momentum of the $\rho$ meson in the nucleon
rest frame. In fact, this is the expression for $ \beta (k_0, {\bf k}) $
normally used in the nuclear matter calculation in the limit of low nuclear
density $\varrho$ or high momentum $ |{\bf k}| \gg p_F $. The rho meson
optical potential obtained from its self-energy, i.e., $ V_{O\rho}
= \frac{1}{2k_0} \Pi_\rho $, is transformed in the proton nucleus cm system
using proper kinematics. The radial density distribution  $\varrho$ for the
$^{12}$C nucleus is expressed in the Eq.~(\ref{dc12}).

Using the $\rho$ meson self-energy described above, we calculate
the mass distribution spectra for the coherent rho meson produced in the
$(p,p^\prime)$ reaction on $^{12}$C nucleus. The beam energy for this
reaction is taken equal to 3.5 GeV, and the ejectile $p^\prime$ emission
angle is taken fixed at $1^0$. In Fig.~1, the cross sections arising due
to $s$ wave $\rho N$ resonances are presented, and those due to $p$
wave $\rho N$ resonances are shown in Fig.~2. It is noticeable in these
figures that the cross sections due to $p$ wave $\rho N$ resonances are
much larger than those due to the $s$ wave $\rho N$ resonances. In
addition, these figures also show that all $s$ wave or $p$ wave $\rho N$
resonances do not contribute identically to the cross section.
It occurs since the cross section
of this reaction depends upon various quantities, such as
coupling constant $f_{RN\rho}$, spin-isospin factor $S_\Sigma$, hadron
parameters for the resonance, .... etc. These quantities, as mentioned
earlier, are listed in Table-1 for all resonances. Remarkably, the
magnitude of the cross section strongly depends on the
resonance($R$)-rho meson($\rho$)-nucleon($N$) coupling constant
$f_{RN\rho}$. As shown in Table-1, $f_{RN\rho}$ for the $p$ wave $\rho N$
resonances are relatively larger. They lie within the range 6.3 to 15.3,
whereas $f_{RN\rho}$ for the $s$ wave $\rho N$ resonances are within the
range of 0.9 to 7.0. Therefore, the cross section due to $p$ wave $\rho N$
resonance is expected to be larger than that due to $s$ wave $\rho N$
resonance.

Fig.~1 shows that the $\rho^0$ meson production cross section due to
$N(1520)$ resonance (solid curve (a)) is the largest compared to those
originating due to other $s$ wave $\rho N$ resonances. This resonance, as
compared in Table-1, has the largest $\rho N$ coupling constant, i.e.,
$ f_{RN\rho} = 7 $. The spin-isospin factor $S_\Sigma$ for it
(equal to 8/3) is also quite significant. The peak cross section due to
this resonance is found equal to 5.64 $\mu$b/(GeV sr), and it appears at
720 MeV in the $\rho$ meson mass distribution spectrum.
The second largest cross section distribution for the rho meson production,
presented by the dot dot dashed curve (b), arises due to $\Delta (1700)$
resonance. This curve shows that the cross section at the peak is equal
to 1.15 $\mu$b/(GeV sr), and the peak position appears at the $\rho$
meson mass equal to 710 MeV. For this resonance, values for $f_{RN\rho}$
and $S_\Sigma$ are equal to 5 and 16/9 respectively.
The $\rho$ meson production cross section due to $\Delta (1620)$ resonance
(dashed curve (c)) is 0.37 $\mu$b/(GeV sr) at the peak, appearing at the
$\rho$ meson mass equal to 710 MeV. $f_{RN\rho}$ and $S_\Sigma$ for
$N(1620)$ are equal to 2.5 and 8/3 respectively. The  $\rho$ meson
production cross section due to $N(1650)$ resonance (dotted curve (d),
not properly visible in Fig.~1) is very small. It happens due to
its weak coupling to the $\rho$ meson and nucleon, i.e., $ f_{RN\rho}
= 0.9 $. However, $S_\Sigma$ for $N(1650)$ is significant (equal to 4).

The rho meson mass distribution spectra due to $p$ wave $\rho N$ resonances,
as mentioned earlier, are presented in Fig.~2.  As shown in this figure, the
maximum cross section arises due to $\Delta (1905)$ resonance (see dashed
curve (a)). For this resonance, $f_{RN\rho}$ and $S_\Sigma$ are equal to
12.2 and 4/5 respectively. The cross section at the peak is equal
to 36.63 $\mu$b/(GeV sr), and it appears at the $\rho$ meson mass equal to
640 MeV. The dot dot dashed curve (b) represents the second largest cross
section distribution originating due to $\Delta (1232)$ isobar. This curve
shows its peak at the $\rho$ meson mass equal to 720 MeV, and the cross
section at the peak is equal to 35.17 $\mu$b/(GeV sr). The constant
$f_{RN\rho}$ for the $\Delta (1232)$ coupling to the $\rho$ meson and
nucleon is quite large. The magnitudes for $f_{RN\rho}$ and $S_\Sigma$ for
this resonance are equal to 15.3 and 16/9 respectively.
It must be mentioned that $\Delta (1232)$ does not have measurable branching
ratio into two pions. Therefore, the value of the $\Delta N\rho$ coupling
constant, i.e., $ f_{RN\rho} (\equiv f_{\Delta N\rho} ) $, can't be
determined from the measured decay width. Since the magnitude of
$f_{\Delta N\rho}$ is uncertain, the cross section arising due to
$\Delta (1232)$ resonance can be interpreted only qualitatively. We have
taken the above value of $f_{\Delta N\rho}$ since it is used by Peters
et al., \cite{ppllm} to calculated the rho meson self-energy.

The $\rho^0$ meson production cross section due to $N(1720)$ resonance
(shown by the solid curve (c)) in Fig.~2 is also significant.
$f_{RN\rho}$ and $S_\Sigma$ for this resonance are 7.8 and 8/3 respectively.
The calculated spectrum due to this resonance shows a peak at the $\rho$
meson mass equal to 720 MeV, and the cross section at the peak is 31
$\mu$b/(GeV sr). The $\rho^0$ meson production cross section due to
nucleon $N(940)$ is presented by the dot dashed curve (d) in Fig.~2.
The $\rho NN$ coupling constant $f_{NN\rho}$ and $S_\Sigma$ are taken
equal to 7.7 and 4 respectively. This curve shows that the cross
section at the peak is equal to 18.99 $\mu$b/(GeV sr) and the peak
position appears at the $\rho$ meson mass equal to 720 MeV. The cross
section due to $N(1680)$ resonance (dotted curve (e)) is least compared
to those due to other $p$ wave $\rho N$ resonances. For this resonance,
$f_{NN\rho}$ and $S_\Sigma$ are equal to 6.3 and 6/5 respectively.

The coherent and incoherent addition of the cross sections due
to all ($s$ and $p$ wave) $\rho N$ resonances have been compared in Fig.~3.
The incoherent addition of the cross sections due to all $\rho N$
resonances (dot dashed curve) shows the cross section at the peak
is equal to $\sim 0.126$ mb/(GeV sr) and it appears at the $\rho$ meson
mass equal to 710 MeV.
The peak cross section due to coherent contribution of all $\rho N$
resonances (solid curve) appears at 730 MeV in the $\rho$ meson mass
distribution spectrum, and its value is equal to 1.35 mb/(GeV sr).
Therefore, the coherent addition
increases the cross section drastically. It enhances the value of the
cross section at the peak by a factor about 10.7 over that due
to incoherent contribution. The peak position due to later appears 20
MeV higher in the $\rho$ meson mass distribution spectrum.

The calculated results presented in Figs.~$1-3$ are based on the choice of
the $\rho$ meson self-energies (given in Eqs~(\ref{ses}) - (\ref{bta}))
evaluated by Peters et al., \cite{ppllm}. As mentioned earlier, this
is a preferable choice in the nuclear matter calculation. However, this
choice had been criticized by Post et al. \cite{plm}, since it leads
serious uncertainty in the non-relativistic calculation.
According to them, the $\rho$ meson self-energies in Eqs.~(\ref{ses}) and
(\ref{sep}) also should be calculated in the resonance rest frame. This
approach gives very reasonable agreement with the relativistic calculation
for the rho meson self-energy \cite{plm}. Therefore, the calculated cross
section using this approach can be considered as the relativistic results.
Since the spectra appearing in Figs.~$1-3$ are due to the non-relativistic
calculation for the $\rho$ meson self-energy, we refer them henceforth
as non-relativistic results. Another advantage of the resonance rest frame
calculation is that the complicated relativistic calculation for the rho
meson self-energy due to spin-$\frac{5}{2}$ $\rho N$ resonance can be
avoided. In the resonance rest frame, the form for the Linhard function
$ \beta (k_0, {\bf k}) $ in Eq.~(\ref{lhfm}) can be simplified to

\begin{equation}
\beta (k_0, {\bf k})
= \frac{1}{2} \varrho
  \frac{ m_R - \frac{i}{2}\Gamma_R - E_N({\bf k}) }
  { k^2_0 - [m_R - \frac{i}{2}\Gamma_R - E_N({\bf k})]^2 },
\label{bta2}
\end{equation}
where $k$ is the four momentum of the $\rho^0$ meson in the resonance
rest frame. All other quantities have been defined earlier.

To explore the relativistic effect on the cross section, we recalculate
the $\rho$ meson mass distribution spectra in the $(p,p^\prime)$ reaction
using the relativistic $\rho$ meson nucleus optical potential
illustrated above. The cross sections originating due to $s$ wave $\rho N$
resonances are shown in Fig.~4. This figure looks qualitatively
similar to the non-relativistic results presented  in Fig.~1.
It is noticeable that magnitudes of the cross sections in the relativistic
calculation have gone up significantly compared to those obtained in the
non-relativistic case. Specifically, the enhancement in the cross section
is found more than 3 times larger for the $N(1520)$ and $\Delta (1700)$
resonances. This enhancement occurs due to the increase in the rho meson
potential, listed in Table-2, in the relativistic calculation.

The cross section due to $N(1520)$ resonance (solid curve (a) in Fig.~4)
is distinctly largest in the relativistic case also. The peak in the
calculated $\rho$ meson mass distribution spectrum due to this resonance
appears at 730 MeV, and the cross section at the peak is equal to 20.04
$\mu$b/(GeV sr). The second largest cross section for the rho meson mass
distribution arises due to the $\Delta (1700)$ resonances (dot dot dashed
curve (b)). The cross section at the peak is equal to 3.82 $\mu$b/(GeV sr),
and the peak appears at the rho meson mass equal to 730 MeV.
The dashed curve (c) represents the mass distribution spectrum for the
rho meson due to the $\Delta (1620)$ resonance. The peak cross section
for this distribution is equal to 0.70 $\mu$b/(GeV sr), and it appears
at the $\rho$ meson mass equal to 720 MeV. Similar to the non-relativistic
case, the cross section due to $N(1650)$ resonance (dotted curve (d),
not prominently seen in Fig.~4) is insignificant compared to those due
to other $s$ wave $\rho N$ resonances.

The calculated (relativistic) rho meson mass distribution spectra due
to $p$ wave $\rho N$ resonances are shown in Fig.~5. Unlike to the case
of $s$ wave $\rho N$ resonances, these spectra are not qualitatively
similar to the non-relativistic results presented in Fig.~2. The $\rho$
meson mass distribution spectra in Fig.~5 show that the largest cross
section arises due to $\Delta (1232)$ resonance. The second largest cross
section originates due to the $\Delta (1905)$ resonance.
It is exactly opposite to the non-relativistic
results (see in Fig.~2). We again mention that the cross section due
to $\Delta (1232)$ resonance can be described only qualitatively, since
uncertainty exists in determining the $\Delta N\rho$ coupling constant
$f_{\Delta N \rho}$. We do not incorporate the cross section due to
$N(940)$ in Fig.~5, since the relativistic calculation for it, as
mentioned by Post et al. \cite{plm}, leads huge over estimation of the
elementary rho production (e.g., $\gamma N \to \rho N$ and
$\pi N \to \rho N$) data.
It is remarkable that the relativistic calculation for the $p$ wave $\rho N$
resonance  gives on an average about $67\%$ less cross section
compare to that due to non-relativistic calculation. This reduction occurs
since the relativistic optical potential due to $p$ wave $\rho N$
resonance-hole polarization, as presented in Table-2, is less than that due
to non-relativistic calculation. This is in contrast to the $s$ wave
$\rho N$ resonance case.

The cross section distribution due to the $\Delta (1905)$ resonance is
described by the dashed curve (a) in Fig.~5. It shows that the peak cross
section is equal to 20.92 $\mu$b/(GeV sr), and the peak position appears at
650 MeV in the $\rho$ meson mass distribution spectrum. The dot dot dashed
curve (b) represents the $\rho$ meson mass distribution spectrum arising due
to the $\Delta (1232)$ resonance. This curve shows that the calculated
cross section at the peak is equal to 26.84 $\mu$b/(GeV sr), and the peak
appears at the rho meson mass equal to 720 MeV.
The solid curve (c) illustrates the $\rho$ meson mass distribution spectrum
due to the $N(1720)$ resonance. The cross section due to it is equal to
19.68 $\mu$b/(GeV sr) at the peak, and it appears at the rho meson mass
equal to 720 MeV. The cross section due to $N(1680)$
resonance (dotted curve (d)) is the smallest amongst those due to other
$p$ wave $\rho N$ resonances. To be mentioned, the non-relativistic
calculation corroborates this finding qualitatively.

The calculated relativistic results for the coherent and incoherent
contributions to the cross section arising from all ($s$ and $p$ wave)
resonances are compared in Fig.~6. The calculated cross
sections in this case show features qualitatively similar to the
non-relativistic results presented in Fig.~3. The incoherent addition of
the cross sections due to all $\rho N$ resonances gives smaller cross
section (dot dashed curve). It is equal to 0.11 mb/(GeV sr) at the peak,
appearing at the $\rho$ meson mass equal to 720 MeV. Whereas, the coherent
contribution of all $\rho N$ resonances (solid curve) enhances the cross
section by a factor about 7.82 to 0.86 mb/(GeV sr). The peak position in
the later case appears at 730 MeV in the $\rho$ meson mass distribution
spectrum.

\section{ CONCLUSIONS }

The $\rho$ meson nucleon resonance dynamics in the nucleus have been
studied through the coherent $\rho^0$ meson production process in the
proton nucleus reaction. This study shows that the cross section
strongly depends upon the $\rho$ meson-nucleon-resonance coupling
constant. It also depends on various other quantities, such as the
spin-isospin transition factor and the hadron parameters for the
resonance .... etc. The cross sections arising due to $p$ wave $\rho N$
resonances are larger than those
due to $s$ wave $\rho N$ resonances. Amongst the $s$ wave resonances, the
largest cross section originates distinctly due to $N(1520)$ resonance.
The relativistic results are qualitatively similar to the
non-relativistic results in the case of $s$ wave $\rho N$ resonances.
The relativistic calculation for these resonances gives larger cross
sections. These features are opposite in the case of $p$ wave $\rho N$
resonances dynamics. The coherent contribution of all resonances enhances
the cross section drastically over that due to incoherent contribution.

\section{ ACKNOWLEDGEMENTS }
The author gratefully acknowledges A.K. Mohanty, R.K. Choudhury
and S. Kailas.

\newpage

\begin{table}[h]
\caption { Values for $\rho N$ resonances' parameters taken from the
Ref.~\cite{ppllm}. $L_{2I,2J}$ is the spectroscopic symbol for the
pion-nucleon resonance state. $S_\Sigma$ denotes the spin-isospin transition
factor. All other symbols carry their usual meanings. }
\vspace{0.5 cm}
\centering
\begin{tabular}{  |c  |c  |c  |c  |c  |c  |c  |c|  }
\cline{1-7}
$R(m_R)L_{2I,2J}$ &       $I(J^P)$               & $l_\rho$ & $\Gamma_\pi$ & $\Gamma_\rho$ & $f_{RN\rho}$ & $S_\Sigma$ \\
\hline
$N(1520)D_{13}$      & $\frac{1}{2}(\frac{3}{2}^-)$ &    0     &      95      &       25      &      7.0     &    8/3     \\

$\Delta(1620)S_{31}$ & $\frac{3}{2}(\frac{1}{2}^-)$ &    0     &     130      &       20      &      2.5     &    8/3     \\

$N(1650)S_{11}$      & $\frac{1}{2}(\frac{1}{2}^-)$ &    0     &     135      &       15      &      0.9     &     4      \\

$\Delta(1700)D_{33}$ & $\frac{3}{2}(\frac{3}{2}^-)$ &    0     &     180      &      120      &      5.0     &   16/9     \\

$N(940)-~~$          & $\frac{1}{2}(\frac{1}{2}^+)$ &    1     &      0       &       0       &      7.7     &     4      \\

$\Delta(1232)P_{33}$ & $\frac{3}{2}(\frac{3}{2}^+)$ &    1     &     120      &       0       &     15.3     &   16/9     \\

$N(1680)F_{15}$      & $\frac{1}{2}(\frac{5}{2}^+)$ &    1     &     118      &       12      &      6.3     &    6/5     \\

$N(1720)P_{13}$      & $\frac{1}{2}(\frac{3}{2}^+)$ &    1     &      50      &      100      &      7.8     &    8/3     \\

$\Delta(1905)F_{35}$ & $\frac{3}{2}(\frac{5}{2}^+)$ &    1     &     140      &      210      &     12.2     &    4/5     \\

\cline{1-7}
\end{tabular}
\end{table}

\begin{table}[h]
\caption { Optical potentials for the $\rho$ meson due to $\rho N$
resonance-hole polarizations in the $^{12}$C nucleus. Potentials are
tabulated below for the $\rho$ meson kinetic energy $T_\rho = 1227.25$ MeV.
This value corresponds to $m_\rho=770$ MeV and $T_{p^\prime}=1500$ MeV. }
\vspace{0.5 cm}
\centering
\begin{tabular}{  |c  |c  |c  |c  |c|  }
\cline{1-4}
$\rho N$ partial wave  & $ N/\Delta $        &  $V_{O\rho}$ (MeV) & $V_{O\rho}$ (MeV) \\
 ($l_\rho$)
                       &                     & (Non-relativistic) & (Relativistic)    \\
\hline
                       &  $N(1520)$          &   -1.94, -11.12    &  -2.24, -23.95    \\

$l_\rho=0$ ($s$ wave)  &  $\Delta(1620)$     &    1.24, -2.66     &   0.99, -4.14     \\

                       &  $N(1650)$          &    0.56, -0.52     &   0.51, -0.65     \\

                       &  $\Delta(1700)$     &   -0.38, -4.88     &  -1.20, -10.01    \\

\hline
                       &  $N(940)$           &   34.46, --        &     --, --        \\

                       &  $\Delta(1232)$     &   -8.57, -27.81    &   -1.40, -28.13   \\
$l_\rho=1$
($p$ wave)             &  $N(1680)$          &   -0.99, -4.72     &   -0.69, -4.89    \\

                       &  $N(1720)$          &   14.11, -33.92    &   6.53, -25.61    \\

                       &  $\Delta(1905)$     &    1.17, -22.74    &  -0.96, -20.01    \\

\cline{1-4}
\end{tabular}
\end{table}

\newpage

{\bf Figure Captions}
\begin{enumerate}

\item
The $\rho$ meson mass distribution spectra due to $s$ wave $\rho N$
resonances have been compared. The non-relativistic $\rho$ meson
self-energy is used in this calculation (see text).

\item
The $\rho$ meson mass distribution spectra due to $p$ wave $\rho N$
resonances have been compared. The non-relativistic $\rho$ meson
self-energy is used in this calculation (see text).

\item
The coherent and incoherent contributions to the cross section arising
from all $\rho N$ resonances are compared. The non-relativistic $\rho$
meson self-energy is used in this calculation (see text).

\item
The $\rho$ meson mass distribution spectra due to $s$ wave $\rho N$
resonances have been compared. The relativistic $\rho$ meson
self-energy is used in this calculation (see text).

\item
The rho meson mass distribution spectra due to $p$ wave $\rho N$ resonances
have been compared. The relativistic $\rho$ meson self-energy is used
in this calculation (see text).

\item
The coherent and incoherent contributions to the cross section arising
from all $\rho N$ resonances are compared. The relativistic $\rho$
meson self-energy is used in this calculation (see text).

\end{enumerate}

\end{document}